\documentclass[letterpaper, 10 pt, conference]{ieeeconf}  

\IEEEoverridecommandlockouts                              
\usepackage{amsmath,amsfonts,amsthm}
\usepackage{graphicx}      
\usepackage{float}
\usepackage{subfig}
\usepackage{xcolor}

 \theoremstyle{definition}
\if@secthm
  \newtheorem{thm}{Theorem}[section]
  \@addtoreset{thm}{section}
\else
  \newtheorem{thm}{Theorem}
\fi
\newtheorem{cor}[thm]{Corollary}

\newtheorem{assum}[thm]{Assumption}
\newtheorem{prop}[thm]{Proposition}
 \theoremstyle{definition}

\newtheorem{rem}[thm]{Remark}

\usepackage[prependcaption,colorinlistoftodos]{todonotes}

\newcommand{\xq}[1]{\textcolor{black}{#1}}
\newcommand{\icl}[1]{\textcolor{black}{#1}}

\newcommand{\ic}[1]{\textcolor{black}{#1}}
\newcommand{\icc}[1]{\textcolor{black}{#1}}
\newcommand{\xqq}[1]{\textcolor{black}{#1}}
\newcommand{\ill}[1]{\textcolor{black}{#1}}
\newcommand{\xxq}[1]{\textcolor{black}{#1}}
\newcommand{\ilc}[1]{\textcolor{black}{#1}}
\newcommand{\ccc}[1]{\textcolor{black}{#1}}

\newcommand{\xqr}[1]{\textcolor{black}{#1}}

\title{\LARGE \bf
Frequency Control and Power Sharing in\\ Combined Heat and Power Networks
}

\author{Xin Qin and Ioannis Lestas
\thanks{X. Qin and I. Lestas are with the Department of Engineering, University of                  Cambridge, Cambridge, CB2 1PZ, UK
        {\tt\small \{xq234, icl20\}@ cam.ac.uk}}%
}

\begin{document}

\maketitle
\thispagestyle{empty}
\pagestyle{empty}

\begin{abstract}

We consider the problem of using district heating systems as ancillary services for primary frequency control in power networks.  We propose a novel power sharing scheme for heating systems based on the average temperature, which enables an optimal power allocation among the diverse heat sources without having a prior knowledge of the disturbances.
We then discuss two approaches for heating systems to contribute to frequency regulation in power networks. We show that both approaches ensure stability in the combined heat and power network and facilitate optimal power allocation among the different energy sources.

\end{abstract}
\section{Introduction}

As decarbonization efforts intensify, the shift towards electrified heating underscores the significance of heat pumps. These high-efficiency devices for converting electricity to heat are gaining popularity with increasing government support.
Unlike traditional electric loads, heat pumps can provide rapid response capabilities and interlink two different energy sectors, the electric power system and the heating system.
The power system requires real-time power balance whereas the temperature dynamics in a heating system have much higher inertia. Thus, if controlled properly, the heating system can provide important support to the frequency control mechanisms in power systems.

However, integrating heat pumps in district heating systems into power system frequency control is a non-trivial problem where various challenges need to be addressed.
The \ccc{dynamics} of primary frequency control raise questions about maintaining stability of the combined heat and power systems amid renewable and load fluctuations. Moreover,  achieving an 
\ccc{appropriate power distribution} 
among the different electric and heat sources is also challenging when heat pumps contribute to frequency regulation. Additionally, \ccc{the combined heat and power network has diverse coupled dynamics associated with the heat and power system that are more challenging to address.}

A number of studies such as the one in \cite{ref_mit2} have contributed to the modeling of heat pumps and heating networks, focusing on the design of control strategies specifically tailored to frequency regulation with a single heat pump. Despite these efforts, the broader system-wide stability has not been extensively explored.
Various control strategies for frequency regulation via heat pumps include those in~\cite{ref_onoff1}, \cite{ref_onoff2}, \cite{ref_pid1}. However, these studies do not provide a stability analysis in a general power network and its coupling with heating networks. Recent advancements, such as the work in \cite{pump_schiffer}, introduce models that incorporate the combined dynamics of heat and power networks. However, the problem of achieving power sharing in heating systems without having a prior knowledge of the disturbances needs to be addressed.
In particular, without such power sharing there is a risk that costly heat sources in a heating network may disproportionately compensate for the heat pump contributions to frequency control, discouraging users from actively participating in frequency regulation efforts.

In this paper, we consider heat pumps that are part of a district heating system and provide an ancillary service to the power grid by contributing to primary frequency regulation. Our contributions can be summarized as follows.
We introduce a novel optimal power sharing scheme among the different heat sources in the heating network. This is based on the use of the average temperature as a control signal and achieves optimal power sharing without having a prior knowledge of the disturbances applied. We also present two schemes for heating systems to contribute to frequency regulation in power networks. When \ccc{these schemes} are implemented, \ccc{it is shown that the} stability of the combined heat and power network is maintained when a general network topology is considered. Furthermore, \ccc{an optimal power sharing can be achieved between diverse electric and heat sources that is quantified within the paper.}

\ccc{The paper is structured as follows. In section \ref{sec:prel} we provide various preliminaries associated with the problem formulation. In section \ref{sec:form} we describe the combined heat and power network that will be considered. In sections \ref{sec:stability} and \ref{sec:sharing} we present the main results associated with stability and power sharing. Extension associated with general passive power and heat generation dynamics are presented in section~\ref{sec.generalpas}. \xqr{Section~\ref{sec.simu} presents the simulation results to demonstrate the effectiveness of the proposed methods.} Finally conclusions are drawn in section \ref{sec:conclusions}.} \xqr{Due to page constraints, the proofs have been omitted.}

\section{Preliminaries}\label{sec:prel}

We use $\textbf{0}_{n}$ and $\textbf{1}_{n}$ to denote the
$n$-dimensional vector of zeros and ones, respectively. For simplicity in the presentation 
we will omit the subscript in the text.
For a vector $x\in \mathbb{R}^n$, we use $diag(x)$ to denote a diagonal matrix with elements $x_i$ in its main diagonal. For vectors $x\in\mathbb{R}^n$, $y\in\mathbb{R}^m$ we use the notation $col(x, y)$ to denote the vector $[x^T \ y^T]^T $ . \xqq{For a function $f(x)$, we use $f'(x) = \frac{df(x)}{dx}$ to denote its first-order derivative. We use $f^{-1}(y)$ to represent the preimage of the point $y$ under the function $f$, i.e., $f^{-1}(y) = \{x:~ f(x) = y\}$. }

We consider in this paper an electric power network at sub-transmission or transmission level and a local heating network coupled together via a heat pump providing ancillary services to the power grid. A graph will be defined for each of these two network systems, respectively, which will then be used to define the underlying physical dynamics in the two networks and the way these are coupled.

In particular, to describe the interconnections in the electric power network we consider a directed graph $(N_e, E_e)$, where $N_e = \left\{1, . . . , \ccc{n_{N_e}} \right\}$ is the set of electric buses and $E_e \subseteq N_e \times N_e$ is set of all power lines.
We use $(i,j)$ to denote the link connecting buses
$i$ and $j$ and assume that the directed graph $(N_e,E_e)$ has an arbitrary orientation, so that if $(i,j)\in E$ then $(j,i)\notin E$.
We use the set $N^G_e = \left\{1, . . . , \ccc{n_{N^G_e}} \right\}$ to denote the set of generators.
We use $i:~i \rightarrow j$ and $k:~j \rightarrow k$ to indicate the predecessors and successors of bus $j$, respectively. 
\ccc{We use the set $H = \left\{1, . . . , \ccc{n_{H}} \right\}$ to denote the set of heat pumps that interconnect the electric and heating systems.}

Similarly, we use a directed graph $G_h = (N_h, E_h)$ to describe the heating network as detailed
in~\cite{ref_machado}, where the direction of the edge is the same as the mass flow. $N_h$ is the set of heat nodes and $E_h$ is the set of heat edges that represent all types of single-input-single-output heat devices including pipelines and heat exchangers. There are three types of heat edges: \ccc{edges associated with} Heat \ccc{pumps \ccc{denoted by set} \xxq{$H_h$},} conventional heat sources like gas boilers \ccc{denoted by set} $E_h^G$, and heat \ccc{loads}, \ccc{denoted by set} $E_h^L$, which use heat exchangers to exchange heat energy with the heating network.
We use the temperature vectors $T^E = col(T^E_{i \in \ccc{H_h}}, T^E_{i \in E_h^G}, T^E_{i \in E_h^L}) $ to denote the edge temperatures of the different type of edges. We use vector $T^N$ to denote the temperature of heat nodes. Then we use vector $T = col(T^E, T^N)$ to aggregate the edge and node temperatures in the heating system.

The following simplifying assumptions are made for the combined heat and power network in the models that will be described in the next section. These are assumptions relevant at \ccc{slower/intermediate} timescales as is the case with the temperature dynamics of \ccc{the heating} systems that will be considered.
Bus voltage magnitudes in the power network are approximately 1 per-unit (p.u.) for each $j\in N_e$.  Lines $(i,j) \in E_e$ are lossless and characterized by their susceptances $B_{ij} = B_{ji} > 0$.
Also reactive power flows do not affect bus voltage angles and frequencies.

\section{Model Formulation}\label{sec:form}
In this section, we describe the dynamic model of the combined heat and power network.

\subsection{Electric power system}
\begin{subequations} \label{eq.edynamics}
We will be using the swing equation to describe the frequency dynamics at each bus $i$ in the electrical power network.
\begin{align}
 M_j \dot \omega_j &=  -p^L_j - p^P_j + p^G_j  - p^U_j  + \sum_{i:~i \rightarrow j} p_{ij} - \sum_{k:~j \rightarrow k} p_{jk}, \nonumber \\
 &~~~~~ \quad \qquad \quad \qquad \qquad  \qquad j \in N_e  \\
 \dot \eta_{ij} &= \omega_i - \omega_j, ~~~~\qquad \qquad \quad ~(i,j) \in E_e \label{eq.line1}\\
 p_{ij} &= B_{ij} \sin(\eta_{ij})-p_{ij}^{\text{nom}}, \quad ~~~(i,j) \in E_e. \label{eq.line2}
\end{align}
\end{subequations}
Variable $\omega_{j}$ is the frequency deviation from a nominal value (50 or 60Hz) at bus $j$. 
Variable $\eta_{ij}$ is the angle difference between bus $i$ and $j$. Variables \ccc{$p^G_j$, $p^U_j$, $p^P_j$, $p^L_j$ represent deviations from nominal values, with $p^G$, $p^U$, $p^P$, $p^L$ denoting corresponding vectors.} 
In particular, $p^G_j$ represents generation at bus $j$. Variable $p^U_j=D_j\omega_j$ where $D_j>0$ is a constant, represents system damping and frequency dependent uncontrollable loads at bus $j$. Variable $p_j^L$ denotes the deviation from a nominal value of a step change in the demand at bus $j$. Constant $M_j>0$ indicates the generator inertia at bus $j$. Constants $B_{ij}>0$ are the line susceptance. Variable $p^P_j$ denotes the electric power consumed by the heat pump at bus $j$. \ccc{Note that this variable is zero if there is no heat pump connected to bus $j$.}

\subsection{Heating system}

In this section, we describe the temperature dynamics of the heating system based on the model in~\cite{ref_machado}. Equation~\eqref{eq.edge_s} quantifies the rate of change of the temperature at an edge which is
dependent on the temperature difference between its inlet and outlet as well as the heat power consumed/generated at this edge. Equation~\eqref{eq.node_s} describes the thermal dynamics at a heat node, indicating that the rate of change of the thermal energy at node $k$ (left hand side) equals its net sum thermal energy input flows 
(right hand side).
\begin{subequations}
\begin{align}
    \rho C_{p} V^E_j \dot{T}^E_{j}&=\rho C_{p} q^E_j \left({T}^N_k-{T}^E_{j}\right) + h^G_{j} + h^P_{j}- h^L_{j}, \nonumber \\
            & \qquad \qquad \qquad j \in E_h, \  k \in N_h \label{eq.edge_s}  \\
    \rho C_{p} V^N_k \dot{T}^N_k &= \sum_{j \in \mathfrak{T}_{k}} \rho C_{p} q^E_j T^E_j-\sum_{j \in \mathfrak{T}_{k}} \rho C_{p} q^E_j {T}^N_k,  \nonumber \\
            & \qquad \qquad \qquad  k \in {N_h}, \label{eq.node_s}
\end{align}
\end{subequations}
where $q^E_j$ is the mass flow of edge $j$. Variable ${T}^E_{j}$ is the temperature of edge $j$. Variables $V^E_j$ and $V^N_k$ are the volumes at edge $j$ and node $k$, respectively. Variables $h^P_{j}$, $h^G_{j}$,  and $h^L_{j}$ are the powers of the heat pump, the conventional heat source,  and the load power at edge $j$, respectively. If edge $j$ is associated with a heat pump, $h^G_j = h^L_j = 0$;
Similarly, if  edge $j$ is a non-pump heat source, $h^P_j = h^L_j = 0$, and if edge $j$ is a heat load, $h^G_j = h^P_j = 0$ \xq{whereas} $h^L_j$ is a given constant. $C_{p}$ and $\rho$ are the capacity and density of water, respectively. In the normalized per-unit system, $\rho C_{p} = 1$. Set $\mathfrak{T}_{k}$ indicates the edges that inject flow to node $k$.

 For convenience in \ic{the 
 analysis}, we write equations~\eqref{eq.edge_s} and~\eqref{eq.node_s} into matrix form:
\begin{align}\label{eq.hdynamics}
V \left[\begin{array}{c}\dot{T}^{{E}} \\ \dot{T}^{N}\end{array}\right]
= - A_h\left[\begin{array}{c}T^{{E}} \\ T^{N}\end{array}\right]+  \left[\begin{array}{c} h^G+h^P-h^L \\ \textbf{0} \end{array}\right] ,
\end{align}
where $h^P$, $h^G$, and $h^L$ are vectors with elements $h^P_j$ , $h^G_j$, and $h^L_j$, respectively.
$V = diag(V^E, V^N)$ indicates the volume matrix.

Note that
\begin{align} \label{eq.A}
    A_h=\left[
\begin{array}{cc}diag(q^E) & -diag(q^E) B_{sh} \\
-B_{th} diag(q^E)              & diag(B_t q^E)\end{array}\right]
\end{align}
is a constant 
matrix 
for a given mass flow vector $q^E$, where
$q^E = col(q^E_j)$. $B_{th} = \frac{1}{2} \left(|B_{h}| + B_{h}\right)$, and $B_{sh} = \frac{1}{2} \left(|B_{h}| - B_{h}\right)$, \ccc{where $B_h$ is the} incidence matrix of the heating system. \icc{It can be shown \ill{that $A_h+A^T_h$ is positive semidefinite with a simple eigenvalue at the origin and $A_h$ satisfies $A_h\textbf{1}=0, \textbf{1}^TA_h=0$} \cite{ref_machado}. }

 \subsection{Generation control} \label{subsec.onoff}

Generators contribute to frequency control in the power grid. We consider heat pumps that operate in a mode that provides an ancillary service to the power grid by contributing to frequency control. The heat pumps are also part of a heating network. In particular, we have the following \ic{frequency} regulation schemes for generators\xqr{~\cite{kasis2016primary}}:
\begin{subequations}\label{eq.control}
\begin{align}
    \dot p^G_j &= -p^G_j - \tilde Q_{e,jj} \omega_j, \quad j \in N_e^G \label{eq.controle}
\end{align}
\ccc{where} 
$\tilde Q_{e,jj} = \frac{1}{Q_{e,jj}}$, \ccc{with} 
scalar $Q_{e,jj}$ \ccc{a positive} cost coefficient reflecting the electric generation cost. For convenience, we define two diagonal matrices $Q_{e} := diag(Q_{e,jj})$ and $\tilde Q_{e} := diag(\tilde Q_{e,jj})$.

We consider also conventional heat sources in the heating system contributing to temperature control via the scheme\xqr{~\cite{pump_schiffer}}:
\begin{align}
    \dot h^G_j &= -h^G_j - \ccc{\tilde Q_{h,jj}} \bar T , \quad j \in E_h^G \label{eq.controlh}
\end{align}
\end{subequations}
\ccc{where} 
\ccc{$\tilde Q_{h,jj} = \frac{1}{Q_{h,jj}}$, \ccc{with} 
scalar $Q_{h,jj}$ \ccc{a positive} constant. We also denote $Q_h=diag(Q_{h,jj})$.}
Variable
\begin{align}
    \bar T := \frac{\textbf{1}^T VT}{\mathcal{V}} \label{eq.avgT}
\end{align}
 is the average temperature of the heating system, which is calculated as a weighted mean of temperature deviations.  The weights used reflect the energy stored in the nodes and edges of the heating system, where $\mathcal{V} = \textbf{1}^T V \textbf{1}$ is a constant equal to
 the sum of the heating system volumes.

\begin{rem}The significance of the use of the average temperature $\bar T$  in the control policies is that it leads to appropriate power sharing properties in the heating system in real time, without having an a prior knowledge of the disturbances applied. This is shown in section \ref{sec:sharing}.
The stability of the combined heat and power network is also shown in section \ref{sec:stability} when these regulation schemes are used, \ccc{in conjunction with frequency support mechanisms by heat pumps described in the next section.}
\end{rem}

\begin{rem}
First order dynamics in~\eqref{eq.control} are used here for simplicity in the presentation. In Section~\ref{sec.generalpas} \ic{we will} generalize \ccc{the} power generation dynamics in~\eqref{eq.control} \icc{to general classes of 
input strictly passive systems} from $\omega_j$ to $p^G_j$ and from $\bar T$ to $h^G_j$.
\end{rem}

\subsection{Heat pump} \label{subsec.pumpm}

When participating 
\ic{in frequency regulation, we assume \ccc{that} the heat pump operates around an operating point where the 
coefficient of} performance (CoP) is approximately a constant~\cite{pump_kim}. The heat pump's electric power consumption and heat power generation are
\begin{align}
    h^P_{j_k} &=  C_o \ccc{p^P_{i_k}}, \quad k \in H \label{eq.pumpmodel}
\end{align}
where $C_o$ indicates the CoP of the heat pump. 
\xxq{\ccc{Note that for each $k\in H$ there are indices $i_k,j_k$, where $i_k \in H_e$ is the bus the heat pump connects to in the electrical network and $j_k \in H_h$ is the corresponding edge in the heating network it is associated with. We denote $H_e$ the set of buses in the electrical network heat pumps are connected to and  $H_h$ the set of edges in the heat network associated with heat pumps.}}

We consider two \xq{schemes} for the heat pump participating in frequency regulation:
\begin{itemize}
    \item \textbf{Mode 1:} Frequency dependent load. In this mode, the heat pump works \ic{as a  frequency} dependent load as shown in~\eqref{eq.pumpm1} \xq{to provide frequency support for the power network}.
    \begin{align}
        p^P_j &=  a_{1,j}\omega_j, \quad j \in \ccc{H_e} \label{eq.pumpm1}
    \end{align}
    \xq{where $a_{1,j}$ is a positive \ccc{constant.} Equation~\eqref{eq.pumpm1} indicates that we regulate the shaft speed of the heat pump in proportion to the bus frequency.}

    \item \textbf{Mode 2:} Converter-linked load. In this mode, \xqq{we consider the output of the heat pump converter as a separate bus with zero inertia, where the frequency is set by the heat pump, and we have power balance considering the power transfers from \ccc{neighbouring buses}}.\\ 
    The \xqq{scheme} is given by equation~\eqref{eq.pumpm2}. In particular, \xqq{the pump converter} interlinks the electric frequency with the thermal temperature, which allows \ic{to achieve appropriate power sharing between the combined} heat and \ic{power network.}
    \begin{subequations}\label{eq.pumpm2}
    \begin{align}
        0 &=  - p^P_j  + \sum_{i:~i \rightarrow j} p_{ij} - \sum_{k:~j \rightarrow k} p_{jk}, \quad \ccc{j \in H_e}  \label{eq.pumpm2_c2}\\
        \omega_{j} &=  m \bar T 
        \label{eq.pumpm2_c1}
    \end{align}
    \end{subequations}
    \xqq{\ccc{Note that} the heat pump power is dependent on the power transfer from other buses as shown in the power balance constraint~\eqref{eq.pumpm2_c2}}. Scalar $m$ is a coefficient indicating the relation between electric frequency and heat average \ccc{temperature.}
\end{itemize}
\begin{rem}
    \xqq{Mode 2 has two major differences from Mode~1. First, as shown} in equation~\eqref{eq.pumpm2_c1} the frequency of the heat pump bus is set directly by the converter, with this then determining the power consumed by the heat pump via the power transfer from neighboring buses (analogous to matching control schemes in hybrid AC/DC networks). This is in contrast to Mode 1 where the heat pump power is set directly as a function of the grid frequency. \xxq{Second, Mode 2 treats the heat pump converter  as a separate bus, which \ccc{implies that} the \ccc{underlying graph} of the electric power system is different from \ccc{that in} Mode 1.}
\end{rem}

\section{Stability of Combined Heat and Power Network}\label{sec:stability}

In this section, we analyze the stability of the combined heat and power system under the two different participation modes of heat pumps described in the previous section, where the heat network also contributes to frequency control via the control policies of the heat pumps.

\begin{assum} \label{assum.equ}
\ccc{The combined heat and power network}
\eqref{eq.edynamics}-\eqref{eq.pumpmodel}, together with~\eqref{eq.pumpm1} or~\eqref{eq.pumpm2}, respectively, admits an equilibrium
point $x^* = (\eta^*, \omega^*, p^{G\ast}, {T}^*, h^{G\ast})$ with $|\eta^*_{ij}| < \frac{\pi}{2}$,  $\forall ij \in E_e$.
\end{assum}
This assumption on $\eta^*_{ij}$ is a security constraint and is ubiquitous in the power network literature.

\begin{thm}[\icc{Stability}] \label{thm:stability}
    Consider the combined heat and power network described by \icc{\eqref{eq.edynamics}-\eqref{eq.pumpmodel}, \eqref{eq.pumpm1} or~\eqref{eq.edynamics}-\eqref{eq.pumpmodel}, \eqref{eq.pumpm2}.} 
    Consider an equilibrium point of this system that satisfies Assumption~\ref{assum.equ}.
    Then, there exists an open neighbourhood of this equilibrium point such that all solutions of the system starting in this region converge to an equilibrium point.
\end{thm}

\section{Power Sharing in the Combined Heat and Power Network}\label{sec:sharing}

In this section, we show that the use of the average temperature in the control policies in the heating system allows to achieve optimal power sharing. In Mode 1, we show that the power sharing is optimal in the electric and the heating system, separately. In Mode 2, we show that the optimal combined electric and heat power sharing is achieved among all electric and heat sources. 

\subsection{Mode 1 } \label{subsec.oproof_m1}

\begin{prop}[\xqq{Optimality under Mode 1}]~\label{thm.opti_sep}
Consider the combined heat and power network described by \eqref{eq.edynamics}-\eqref{eq.pumpmodel} \xqq{and~\eqref{eq.pumpm1}.}
\ilc{The deviation in the power generation $p^G$ and of the heat pump load $p^P$ at equilibrium after a step change in electrical load $p^L$ is the solution to} 
the optimization \ilc{problem}
\begin{subequations}~\label{eq.opti_sepe}
\begin{gather}
    \min_{p^G, \xqq{p^P, p^U}} C_{1,e} = \frac{1}{2} (p^G)^T Q_e p^G   \nonumber\\
        +\xxq{\frac{1}{2} (p^P)^T Q_p (p^P)^T + \frac{1}{2} (p^U)^T Q_u p^U},
\end{gather}
subject to
\begin{gather}
    \textbf{1}^T p^G = \textbf{1}^T p^L + \textbf{1}^T p^P + \textbf{1}^T p^U,  \label{eq.opti_ec1}
\end{gather}
\end{subequations}
where 
\xxq{\ccc{$Q_p = diag(Q_{p,jj})$ \ilc{with} 
$Q_{p,jj} = \frac{1}{a_{1,j}}$, \ccc{$j\in H_e$} 
and $Q_u = diag(Q_{u,jj})$ 
with $Q_{u,jj} = \frac{1}{D_{j}}$, $j\in N_e$.}}\\
Given the 
\ilc{heat pump power consumption
$\bar p^P$ in the solution of \eqref{eq.opti_sepe}, the deviation in power \ilc{$h^G$} of the conventional heat sources 
is the solution} to the optimization \ilc{problem} 
\begin{subequations}~\label{eq.opti_seph}
\begin{gather}
    \min_{h^G} C_{1,h} = \frac{1}{2} (h^G)^T {Q}_h h^G,
\end{gather}
subject to
\begin{gather}
    \textbf{1}^T h^G = \textbf{1}^T (h^L - h^P).  \label{eq.opti_hc1}
\end{gather}
\end{subequations}
\end{prop}

\subsection{Mode 2} \label{subsec.oproof_m2}

\begin{thm}[\xqq{Optimality under Mode 2}]~\label{thm.opti_joint}
Consider the combined heat and power network described by \eqref{eq.edynamics}-\eqref{eq.pumpmodel} and~\eqref{eq.pumpm2}. Then the power contribution of the electric generators \ilc{$p^G$} and conventional heat sources \ilc{$h^G$} \ilc{
is the solution} to the combined heat and power optimization problem

\begin{subequations}~\label{eq.opti_joint}
\begin{gather}
    \min_{p^G, h^G, \xqq{p^U}} C_2 =\frac{1}{2} (p^G)^T Q_e p^G + \frac{1}{2} (h^G)^T \ccc{\frac{m}{C_o}Q_h} h^G \nonumber \\
    + \xxq{ \frac{1}{2} (p^U)^T Q_u p^U},
\end{gather}
subject to
\begin{align}
    \textbf{1}^T p^G &= \textbf{1}^T p^L + \textbf{1}^T p^P + \textbf{1}^Tp^U ,\label{eq.opti_jointc1}\\
    \textbf{1}^T h^G &= \textbf{1}^T h^L -  \textbf{1}^T h^P,   \label{eq.opti_jointc2} \\
  h^P_{j_k} &=  \xxq{C_o p^P_{i_k}, \quad k \in H}. \label{eq.opti_jointc3}
\end{align}
\ccc{where $Q_u$ is as defined in Proposition \ref{thm.opti_sep}.}
\end{subequations}
\end{thm}

\section{General passive dynamics} \label{sec.generalpas}
\subsection{\xqr{Problem Statement}}
\xq{In practical combined heat and power networks, the \ccc{dynamics of} generators and conventional heat sources are diverse, extending beyond simple first-order dynamics \ilc{as in \eqref{eq.control}.} To address this practical \ilc{issue,} we \ilc{describe in this section how the approach in the paper can} 
accommodate \ilc{general higher order generation} dynamics characterized by
\icc{\ilc{input strictly}} passive systems, \icc{relating $\omega_j$} to $p^G_j$ \icc{and}  $\bar T$ to $h^G_j$, \ilc{respectively}.}

\ilc{More precisely, in the power network we consider $p^G_j$ now as} the output of a nonlinear system \ilc{with input $\omega_j$,} namely
\begin{subequations}\label{eq.epas}
    \begin{align}
    \dot x^s_{e,j} &= -f_{e,j}(x^s_{e,j},~\omega_j), \quad j \in N_e^G,   \\
             p^G_j &= -g_{e,j}(x^s_{e,j},~\omega_j), \quad j \in N_e^G.
\end{align}
\end{subequations}
where $x^s_{e,j}$ is the state of the system, and $f_{e,j}$ and $g_{e,j}$ are \icc{locally} Lipschitz \ilc{functions.} System~\eqref{eq.epas} is said to be locally input strictly passive around the \ilc{equilibrium point}  $\omega_j^*$, $x^{s*}_{e,j}$, if there exist open neighborhoods $\Omega_{j}$ of $\omega_j^*$ and $X^s_{e,j}$ of $x^{s*}_{e,j}$ and a continuously differentiable positive semi-definite function $V^s_{e,j}(x^s_{e,j})$, such that for all $\omega_j \in \Omega_{j}$ and all $x^s_{e,j} \in X^s_{e,j}$,~$j \in N^G_e$, one has
\begin{align}
    \dot V_{e,j}^g(x^s_{e,j}) \leq (-\omega_j + \omega_j^* )(p^G_j - p^{G*}_j) - \phi_{e,j}(-\omega_j + \omega_j^*), \nonumber
\end{align}
where $\phi_{e,j}$ is a \ilc{positive definite} function. At the equilibrium \ccc{point}, we define
\begin{align}
    p^{G*}_j& = K_{p^G_j}(-\omega_j^*),
\end{align}
where $K_{p^G_j}(-\omega_j^*)$ is a static input-output characteristic map, \ilc{which is defined under the assumption that the equilibrium value $p^{G\ast}_j$ is unique for a given constant input $\omega_j^\ast$. We also assume that the equilibrium point $x^{s*}_{e,j}$ is asymptotically stable for subsystem \eqref{eq.epas} for  constant input $\omega_j^*$.}

Similarly, \ilc{we consider} a general class of generation dynamics of conventional heat sources, \ilc{where} we define $h^G_j$ is the output of a nonlinear \ilc{system}
\begin{subequations}\label{eq.hpas}
    \begin{align}
    \dot x^s_{h,j} &= -f_{h,j}(x^s_{h,j},~\bar T), \quad j \in E^G_h,  \\
             h^G_j &= -g_{h,j}(x^s_{h,j},~\bar T), \quad j \in E^G_h.
\end{align}
\end{subequations}
where $x^s_{h,j}$ is the state of the system, and $f_{h,j}$ and $g_{h,j}$ are \ilc{locally} Lipschitz. System~\eqref{eq.hpas} is said to be locally input strictly passive around the equilibrium \ilc{point $\bar T^*$, $x^{s*}_{h,j}$}, if there exist open neighborhoods \ilc{$\mathcal{T}_{j}$ of} $\bar T^*$ and  $X^s_{h,j}$ of $x^{s*}_{h,j}$ and a continuously differentiable positive semi-definite function $V^s_{h,j}(x^s_{h,j})$, such that for all \ilc{$\bar T \in \mathcal{T}_{j}$} and all $x^s_{h,j} \in X^s_{h,j}$,~$j \in E^G_h$, one has
\begin{align}
    \dot V_{h,j}^g(x^s_{h,j})& \leq (-\bar T + \bar T^* )(h^G_j - h^{G*}_j) - \phi_{h,j}(-\bar T + \bar T^*),
\end{align}
where $\phi_{h,j}$ is a \ilc{positive definite} function. At the equilibrium \ilc{point}, we \ilc{similarly} define
\begin{align}
    h^{G*}_j& = K_{h^G_j}(-\bar T^*),
\end{align}
where $K_{h^G_j}(-\bar T^*)$ is a static input-output characteristic map \ilc{and assume  $x^{s*}_{h,j}$ is asymptotically stable for subsystem \eqref{eq.hpas}.}

\begin{assum}~\label{assum.equ_pa}
   \xqq{ Each of the systems defined in~\eqref{eq.epas} and \eqref{eq.hpas} with given inputs and outputs respectively are locally input strictly passive about 
   equilibrium values \ilc{associated with an equilibrium point of the combined heat and power network. The corresponding storage functions also have strict local minima at this equilibrium point.}}
   \end{assum}

\subsection{Stability under General Passive \xqr{Dynamics}}
\begin{cor}\label{cor.sgeneral}
    Consider the combined heat and power network described by \eqref{eq.edynamics}-\eqref{eq.A}, \xqq{\eqref{eq.pumpmodel}, \eqref{eq.epas}, \eqref{eq.hpas}, \eqref{eq.pumpm1} or \eqref{eq.edynamics}-\eqref{eq.A}, \eqref{eq.pumpmodel}, \eqref{eq.epas}, \eqref{eq.hpas}, \eqref{eq.pumpm2}.} Consider \ccc{an equilibrium} point of this system that satisfies Assumptions~\ref{assum.equ} and~\ref{assum.equ_pa}.
    Then there exists an open neighbourhood of this equilibrium point such that all solutions of the system starting in this region converge to an equilibrium point.
\end{cor}

\subsection{Power Sharing under \ccc{General Passive Dynamics}}
\ilc{It can be shown that analogous power sharing results to those stated in Proposition \ref{thm.opti_sep} and Proposition \ref{thm.opti_joint} can be obtained when the generalized generation dynamics \eqref{eq.epas}, \eqref{eq.hpas} are considered. In particular, Propositions \ref{thm.opti_sep}, \ref{thm.opti_joint} hold but with the quadratic terms $\frac{1}{2} (p^G)^T Q_e p^G$, $\frac{1}{2} (h^G)^T \frac{m}{C_o}Q_h h^G$ replaced by $\hat C_{e}(p^G)$, $\frac{m}{C_o}\hat C_{h}(h^G)$ respectively, where}
\begin{subequations}
    \begin{align}
        K_{p^G_j}(\omega^*) = (\hat C^{'}_{e,jj})^{-1}(\xqq{\omega^*}),~j \in N_e^G \\
        K_{h^G_j}(\bar T^*) = (\hat C^{'}_{h,jj})^{-1}(\xqq{\bar T^*}),~j \in E_h^G
    \end{align}
    \end{subequations}
and   $\hat C_{e}(p^G)$, $\hat C_{h}(h^G)$ are assumed to be strictly convex.
This follows from the Karush–Kuhn–Tucker (KKT) conditions for optimality.

\section{Simulation Results} \label{sec.simu}

\begin{figure}[h]%
        \centering
            \includegraphics[trim = 0mm 0mm 0mm 0mm, clip, width = .99\columnwidth]{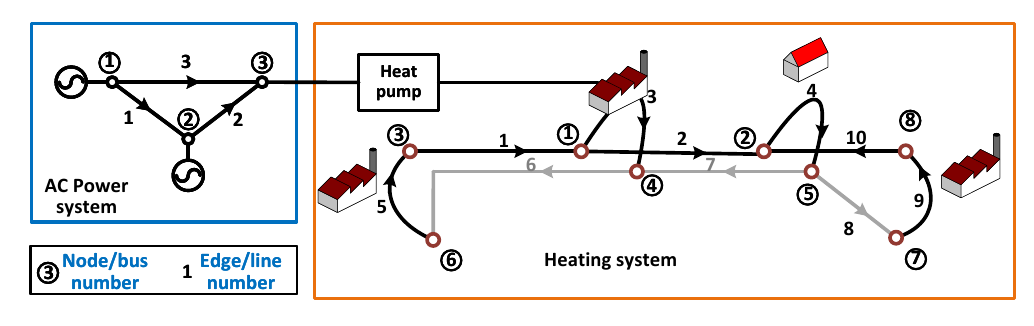}
      \caption{\xqr{Typology of the combined heat and power system.}}
      \label{fig0}
    \end{figure}

We use a \icl{3-bus 
power system} with 2 \icl{generators} and \icl{a 10-edge} heating system with 3 heat sources, \xqr{as shown in Figure~\ref{fig0}}, to demonstrate the effectiveness of the proposed methods. The power network has a \icl{step change in load} of 0.1 p.u. at bus 2 \icl{at} $t=5$s.

\begin{figure}[H]%
	\centering
	\subfloat[~]{
		\includegraphics[trim = 0mm 0mm 0mm 0mm, clip, width = .479\columnwidth]{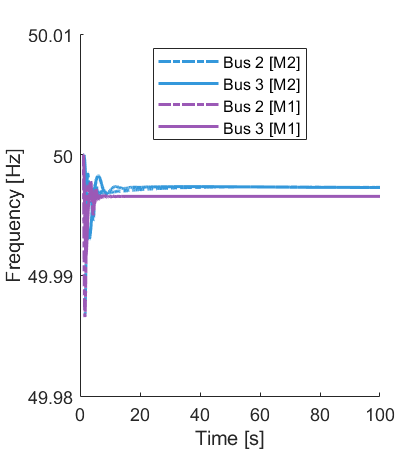}\label{fig1}
	}
 	\subfloat[~]{
		\includegraphics[trim = 0mm 0mm 0mm 0mm, clip, width = .479\columnwidth]{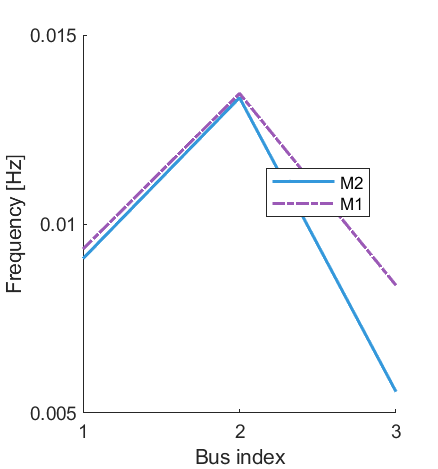}\label{fig2}%
	} 
  \caption{\xqr{(a) Frequency at bus 2 and bus 3, and (b) maximum frequency deviation in the transient response, where M2 denotes Mode 2 and M1 denotes Mode 1.}}
    \label{fig11}
\end{figure}

\xqr{As shown in Figure~\ref{fig1}, both Mode 1 and Mode 2 can stabilize the frequency in the combined heat and power system, while Mode 2 results in lower frequency deviation from the nominal value 50 Hz. 
We also observe that Mode 2 took longer to stabilize the frequency, taking 9.45 seconds compared to 4.7 seconds for Mode 1 at bus 3 as shown in Fig.~\ref{fig1}. Furthermore, we see from Figure~\ref{fig2} that Mode 2 has lower frequency deviation within the transient response, which indicates Mode 2 can better leverage heat inertia to contribute to reducing frequency fluctuations.
}

\begin{figure}[H]%
	\centering
 	\subfloat[~]{
		\includegraphics[trim = 0mm 0mm 0mm 0mm, clip, width = .479\columnwidth]{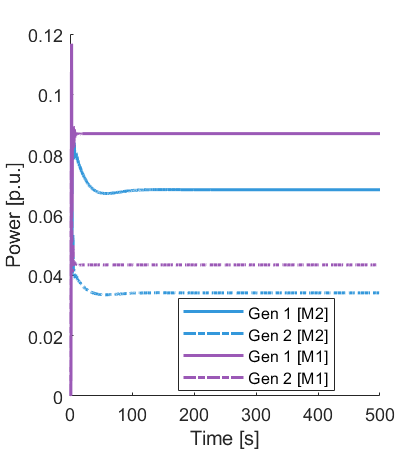}\label{fig3}
	}
 	\subfloat[~]{
		\includegraphics[trim = 0mm 0mm 0mm 0mm, clip, width = .479\columnwidth]{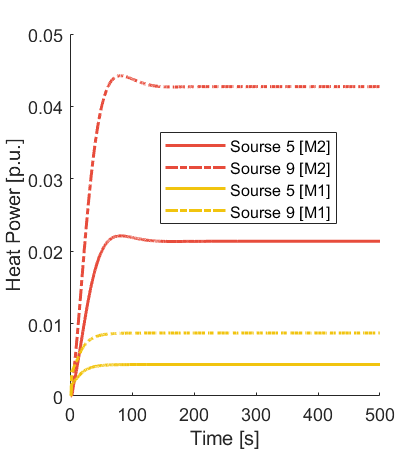}\label{fig4}%
	} 
  \caption{{Power adjustments of (a) generators and  
    (b) heat sources. The cost coefficients of the generators at buses 1 and 2 are 2 and 1, respectively, and the cost coefficients of the heat sources at edges 5 and 9 are 0.5 and 1.}
    }
    \label{fig22}
\end{figure}

\xqr{Figure~\ref{fig22} shows that both Modes 1 and 2 achieve optimal power sharing, as stated in Proposition~\ref{thm.opti_sep} and Theorem~\ref{thm.opti_joint}, respectively. The key difference between the two modes lies in the power distribution: Mode 2 utilizes the heat pump more effectively than Mode 1 for frequency regulation, which reduces the reliance on generators and overall costs. This is because Mode 1 achieves power sharing within each individual system, while Mode 2 extends this optimization to the combined heat and power system.}

\xqr{Based on the simulation result of primary control, if secondary frequency control is present in the power network, the contribution of the heat pump and other heating sources is transient and is still important as it can reduce the frequency fluctuation and the deviation from the nominal value.}

\section{Conclusion}\label{sec:conclusions}
We have shown that the use of the average temperature as a control signal in a district heating system can lead to appropriate optimal power sharing, while maintaining stability guarantees in general network topologies and general generation dynamics. We have also proposed two schemes through which heat pumps can contribute to frequency control with the combined heat and power network maintaining its stability \ccc{and also achieving appropriate power sharing.}

\bibliographystyle{IEEEtran}
\bibliography{main}             
\clearpage
\newpage

\end{document}